\PassOptionsToPackage{unicode}{hyperref}
\PassOptionsToPackage{hyphens}{url}
\PassOptionsToPackage{dvipsnames,svgnames,x11names}{xcolor}
\documentclass[
  12pt]{article}

\usepackage{amsmath,amssymb}
\usepackage{iftex}
\ifPDFTeX
  \usepackage[T1]{fontenc}
  \usepackage[utf8]{inputenc}
  \usepackage{textcomp} % provide euro and other symbols
\else % if luatex or xetex
  \usepackage{unicode-math}
  \defaultfontfeatures{Scale=MatchLowercase}
  \defaultfontfeatures[\rmfamily]{Ligatures=TeX,Scale=1}
\fi
\usepackage{lmodern}
\ifPDFTeX\else
\fi
\IfFileExists{upquote.sty}{\usepackage{upquote}}{}
\IfFileExists{microtype.sty}{% use microtype if available
  \usepackage[]{microtype}
  \UseMicrotypeSet[protrusion]{basicmath} % disable protrusion for tt fonts
}{}
\makeatletter
\@ifundefined{KOMAClassName}{% if non-KOMA class
  \IfFileExists{parskip.sty}{%
    \usepackage{parskip}
  }{% else
    \setlength{\parindent}{0pt}
    \setlength{\parskip}{6pt plus 2pt minus 1pt}}
}{% if KOMA class
  \KOMAoptions{parskip=half}}
\makeatother
\usepackage{xcolor}
\makeatletter
\ifx\paragraph\undefined\else
  \let\oldparagraph\paragraph
  \renewcommand{\paragraph}{
    \@ifstar
      \xxxParagraphStar
      \xxxParagraphNoStar
  }
  \newcommand{\xxxParagraphStar}[1]{\oldparagraph*{#1}\mbox{}}
  \newcommand{\xxxParagraphNoStar}[1]{\oldparagraph{#1}\mbox{}}
\fi
\ifx\subparagraph\undefined\else
  \let\oldsubparagraph\subparagraph
  \renewcommand{\subparagraph}{
    \@ifstar
      \xxxSubParagraphStar
      \xxxSubParagraphNoStar
  }
  \newcommand{\xxxSubParagraphStar}[1]{\oldsubparagraph*{#1}\mbox{}}
  \newcommand{\xxxSubParagraphNoStar}[1]{\oldsubparagraph{#1}\mbox{}}
\fi
\makeatother

\usepackage{longtable,booktabs,array}
\usepackage{calc} % for calculating minipage widths
\usepackage{etoolbox}
\makeatletter
\patchcmd\longtable{\par}{\if@noskipsec\mbox{}\fi\par}{}{}
\makeatother
\IfFileExists{footnotehyper.sty}{\usepackage{footnotehyper}}{\usepackage{footnote}}
\makesavenoteenv{longtable}
\usepackage{graphicx}
\makeatletter
\def\maxwidth{\ifdim\Gin@nat@width>\linewidth\linewidth\else\Gin@nat@width\fi}
\def\maxheight{\ifdim\Gin@nat@height>\textheight\textheight\else\Gin@nat@height\fi}
\makeatother
\setkeys{Gin}{width=\maxwidth,height=\maxheight,keepaspectratio}
\makeatletter
\def\fps@figure{htbp}
\makeatother

\makeatletter
\@ifpackageloaded{caption}{}{\usepackage{caption}}
\AtBeginDocument{%
\ifdefined\contentsname
  \renewcommand*\contentsname{Table of contents}
\else
  \newcommand\contentsname{Table of contents}
\fi
\ifdefined\listfigurename
  \renewcommand*\listfigurename{List of Figures}
\else
  \newcommand\listfigurename{List of Figures}
\fi
\ifdefined\listtablename
  \renewcommand*\listtablename{List of Tables}
\else
  \newcommand\listtablename{List of Tables}
\fi
\ifdefined\figurename
  \renewcommand*\figurename{Figure}
\else
  \newcommand\figurename{Figure}
\fi
\ifdefined\tablename
  \renewcommand*\tablename{Table}
\else
  \newcommand\tablename{Table}
\fi
}
\@ifpackageloaded{float}{}{\usepackage{float}}
\floatstyle{ruled}
\@ifundefined{c@chapter}{\newfloat{codelisting}{h}{lop}}{\newfloat{codelisting}{h}{lop}[chapter]}
\floatname{codelisting}{Listing}

\makeatother
\makeatletter
\@ifpackageloaded{caption}{}{\usepackage{caption}}
\@ifpackageloaded{subcaption}{}{\usepackage{subcaption}}
\makeatother

\ifLuaTeX
  \usepackage{selnolig}  % disable illegal ligatures
\fi
\usepackage[]{natbib}
\usepackage{bookmark}

\IfFileExists{xurl.sty}{\usepackage{xurl}}{} % add URL line breaks if available
\hypersetup{
  pdftitle={Do the Numbers Match the Plot? Statistical-Visual Verification for Automated Exploratory Data Analysis},
  pdfauthor={Zixun FU; Jian Huang; Maojun Sun; Yancheng Yuan},
  pdfkeywords={automated exploratory data analysis, data agents, statistical graphics, statistical-visual verification},
  colorlinks=true,
  linkcolor={blue},
  filecolor={Maroon},
  citecolor={Blue},
  urlcolor={Blue},
  pdfcreator={LaTeX via pandoc}}

\newcommand{\anon}{1}

\begin{document}

\def\spacingset#1{\renewcommand{\baselinestretch}%
{#1}\small\normalsize} \spacingset{1}

%%%%%%%%%%%%%%%%%%%%%%%%%%%%%%%%%%%%%%%%%%%%%%%%%%%%%%%%%%%%%%%%%%%%%%%%%%%%%%

\if1\anon
{
  \title{\bf Do the Numbers Match the Plot? Statistical--Visual Verification for Automated Exploratory Data Analysis}
  \author{Zixun FU$^{1,\dagger}$,
  Jian Huang$^{1,2,\dagger,*}$,
  Maojun Sun$^{2,\dagger}$, and
  Yancheng Yuan$^{1,\dagger,*}$\\[0.4em]
  \footnotesize $^{1}$Department of Applied Mathematics, The Hong Kong Polytechnic University,\\
  \footnotesize Hong Kong SAR, China\\
  \footnotesize $^{2}$Department of Data Science and Artificial Intelligence,\\
  \footnotesize The Hong Kong Polytechnic University, Hong Kong SAR, China\\[0.25em]
  \footnotesize $^{\dagger}$All authors contributed equally to this work. Authors are listed alphabetically by surname.\\
  \footnotesize $^{*}$Corresponding authors: Jian Huang and Yancheng Yuan.\\
  \footnotesize Zixun FU: \href{mailto:zixun.fu@connect.polyu.hk}{\nolinkurl{zixun.fu@connect.polyu.hk}};
  Jian Huang: \href{mailto:j.huang@polyu.edu.hk}{\nolinkurl{j.huang@polyu.edu.hk}}.\\
  \footnotesize Maojun Sun: \href{mailto:mj.sun@connect.polyu.hk}{\nolinkurl{mj.sun@connect.polyu.hk}};
  Yancheng Yuan: \href{mailto:yancheng.yuan@polyu.edu.hk}{\nolinkurl{yancheng.yuan@polyu.edu.hk}}.}
  \date{}
  \maketitle
  \vspace{-2em}
} \fi

\if0\anon
{
  \bigskip
  \bigskip
  \bigskip
  \begin{center}
    {\LARGE\bf Do the Numbers Match the Plot? Statistical--Visual Verification for Automated Exploratory Data Analysis}
  \end{center}
  \medskip
} \fi

\bigskip

\begin{abstract}
Automated exploratory data analysis (EDA) systems can execute statistics correctly while producing misleading interpretations. For example, correlations may be driven by influential observations, and mean differences may arise from skewness, overlap, or sparse subgroups rather than broad distributional shifts. We present EDAgent, a sequential framework that performs targeted visual verification before report synthesis. For each finding, the system records the statistical result, generates a visualization request identifying relevant distributional features, creates the corresponding chart, and uses the observed pattern to retain, qualify, or reconsider the interpretation. We call this process statistical--visual verification. EDAgent includes VisPlan, an 8B-parameter model that converts visualization rationales into detailed chart plans. We evaluate 87 end-to-end reports on EDA-Bench using a six-dimension rubric and GPT-5.1 multimodal evaluation. Maximum-score proportions reached 87.4\% for factuality and visual consistency, 93.1\% for statistical--visual cross-validation, and 96.6\% for analytical narrative. On 166 open-ended VisEval tasks, VisPlan improved the mean overall score from 4.55 (Qwen3-8B) to 4.62, with the largest gain in aesthetics. These descriptive results indicate that visual verification can make automated EDA reports more transparent about how statistical findings relate to underlying data structure.
\end{abstract}

\noindent%
\vfill

\newpage
\spacingset{1.8} % DON'T change the spacing!

\section{Introduction}

An automated exploratory data analysis (EDA) system can report the right number and still give a misleading impression of the data. For example, a Pearson correlation may be numerically correct but dominated by one high-leverage observation, while a difference in group means may be driven by an upper tail rather than a broad distributional shift. Anscombe's Quartet is the canonical reminder that nearly identical summaries can coexist with markedly different data geometry \citep{anscombe1973graphs}. Numerical summaries are therefore not unreliable in themselves; rather, their meaning depends on distributional structure that a small set of aggregates cannot preserve.

This interplay between numerical summaries and data geometry is central to EDA. Statistical graphics are used to inspect data quality, distributions, relationships, and model assumptions before confirmatory analysis \citep{tukey1977exploratory,cleveland1993visualizing,wilkinson2005grammar,donoho201750}. The process is iterative: a statistical finding motivates a plot, the plot changes or sharpens the interpretation, and the revised interpretation motivates a more focused calculation. Recent work on projection pursuit for large datasets provides one example of computation and visualization being developed as complementary analytical operations \citep{duan2025projection}. As more analysis is delegated to large language model (LLM) agents, preserving this interaction becomes a methodological problem rather than merely a matter of report design.

Data agents can translate natural-language questions into code, execute analyses, generate charts, and synthesize reports \citep{sun2025survey,sun2026lambda,hong2025data}. Execution grounding is essential because it ties reported values to program output, and recent benchmarks accordingly assess execution, factuality, and analysis quality \citep{sun2026dsaeval}. Yet successful execution does not ensure that an aggregate statistic is a representative description of the underlying data. We use the term \emph{summary--structure mismatch} for a situation in which a statistic is computed correctly but, when viewed alone, conveys a misleading impression because it suppresses consequential structure. Skewness, influential observations, latent subgroups, nonlinear relationships, and sparse cells can all produce such a mismatch. Automated EDA therefore needs to establish both what was computed and whether important features of the data alter how that result should be interpreted.

Statistical graphics research provides a precedent for treating a plot as evidence rather than decoration. For instance, the lineup protocol places an observed plot among plots generated under a null model and calibrates visual pattern detection against sampling variation \citep{buja2009statistical}. Recent studies have applied this logic to residual diagnostics and change-point perception \citep{li2024plot,fudolig2025change}. Such formal visual-inference procedures specify a null mechanism, an inferential target, and an observer protocol. Our setting is exploratory: a targeted chart is used to examine whether distributional structure changes the interpretation of an executed result, without assigning a visual $p$-value or validating a statistical model.

Natural language-to-visualization (NL2Vis) research provides the technical basis for generating such charts. Cross-domain datasets and systems translate explicit plotting requests into specifications or executable visualizations \citep{luo2021nvbench,maddigan2023chat2vis,wu2024automated}. LIDA decomposes visualization generation into summarization, goal exploration, code generation, and rendering \citep{dibia2023lida}; MatPlotAgent adds query expansion, code debugging, and visual feedback \citep{yang2024matplotagent}; and Data Formulator, DeepVis, and nvAgent improve concept-driven, automated, or collaborative chart authoring \citep{wang2024dataformulator,shuai2025deepvis,ouyang2025nvagent}. These systems typically begin with a requested chart or visualization intent. The complementary problem studied here is to derive a visual verification request from a statistical finding, generate the required view, and return the observed pattern to the reasoning process before the report is written.

Open-ended visualization also raises an evaluation problem because no unique reference chart exists. Quality depends on semantic alignment, data encoding, readability, effectiveness, and design, and a system can succeed on some dimensions while failing on others. VisEval offers a multidimensional benchmark for this setting \citep{chen2024viseval}, while multimodal models make scalable image-level assessment possible \citep{yang2024matplotagent,sun2026dsaeval}. Because model judges can share preferences with the evaluated system, overlook errors not visible in a rendering, and vary across runs, we treat their scores as descriptive measurements rather than independent ground truth.

We propose EDAgent to operationalize \emph{statistical--visual verification}. Given a dataset and a natural-language question, the system (i) executes descriptive-analysis code and extracts statistical findings; (ii) converts a potential vulnerability of each material finding into a visual verification request; (iii) generates targeted charts through a hybrid visualization engine; and (iv) compares the visible structure with the finding that motivated the chart. The resulting verification record states whether the original interpretation can be retained, should be qualified, or needs to be reconsidered. In this workflow, execution establishes the reported value, whereas visualization examines whether that value provides an adequate summary of the relevant data structure.

We evaluate complete reports on EDA-Bench, a suite of 87 public tabular datasets, and examine the visualization-planning component on 166 open-ended tasks adapted from VisEval \citep{chen2024viseval}. EDA-Bench reports are scored on factual consistency, data caveats, statistical--visual cross-validation, analytical narrative, visual contribution, and next-step quality. Because EDA-Bench has no directly evaluated end-to-end comparator and each report receives one judgment from a single multimodal model, its results characterize performance under the adopted rubric rather than comparative superiority or human-validated correctness.

This work makes three contributions. First, it specifies a traceable statistical--visual verification workflow for automated EDA, linking each material interpretation to executed output, a targeted visual verification request, an observed feature, and revised wording. Second, it implements the workflow in EDAgent and develops VisPlan, an 8B-parameter visualization planner that converts analytical rationales into detailed chart plans. VisPlan is an implementation contribution that strengthens the visualization stage, although statistical--visual verification does not depend on this particular planner. Third, it contributes EDA-Bench and an open-ended adaptation of VisEval, together with descriptive results and case studies that demonstrate both the practical value and the current limits of model-based evaluation.

\section{EDAgent: Statistical--Visual Verification}

Given a collection of tabular datasets $\mathcal{D}=\{D_1,D_2,\ldots,D_n\}$ and a natural-language query $Q$, EDAgent produces an evidence-grounded exploratory analysis report $R$. The framework is organized around three sequential components: an execution-grounded statistical stream, a hybrid visualization engine, and a multimodal verification and report-synthesis stage.

\subsection{Scope and Verification Record}

The unit of analysis is the interpretation attached to a material statistical finding, not the predictive performance of a fitted model. For finding $j$, EDAgent constructs a verification record
\[
\mathcal{E}_j=(I_{\mathrm{stat},j},V_{\mathrm{req},j},\mathcal{P}_j,F_j,O_j,R_j),
\]
where $I_{\mathrm{stat},j}$ is the executed result; $V_{\mathrm{req},j}$ identifies a distributional feature that could affect its interpretation; $\mathcal{P}_j$ contains the rendered plots; $F_j$ records the specific visible feature used in the comparison; $O_j\in\{\textit{retain},\textit{qualify},\textit{reconsider}\}$ is the verification outcome; and $R_j$ is the wording admitted to the final report. A record is complete only when the plotted data, visible feature, and revised wording can all be traced back to the executed result.

\begin{table*}[t]
\centering
\small
\setlength{\tabcolsep}{3pt}
\begin{tabular}{p{0.14\textwidth} p{0.22\textwidth} p{0.31\textwidth} p{0.23\textwidth}}
\toprule
Record & Question answered & Required content & Effect on the report \\
\midrule
Executed result $I_{\mathrm{stat},j}$ & What was computed? & Quantity, variables, subset, units, and the program output from which the value was taken. & Establishes the statistical finding that may enter the report. \\
Verification request $V_{\mathrm{req},j}$ & What could make the interpretation too broad? & A named vulnerability such as leverage, skewness, nonlinearity, subgroup heterogeneity, multimodality, or sparse cells. & Determines the data view and graphical encoding required for the check. \\
Rendered check $(\mathcal{P}_j,F_j)$ & What feature is actually visible? & Figure identifier, plotted subset, encoding, and the visible feature used as evidence. & Prevents a chart from being cited without stating what it contributes. \\
Outcome and revision $(O_j,R_j)$ & What wording survives the check? & Retain, qualify, or reconsider, followed by the resulting interpretation. & Controls the language admitted to final synthesis. \\
\bottomrule
\end{tabular}
\caption{Finding-level verification record. Every reported interpretation retains a traceable path from executed output to a targeted visual check and revised wording.}
\label{tab:verification_record}
\end{table*}

The comparison between a rendered chart and its motivating statistical finding leads to one of three practical outcomes. When the visible pattern agrees with both the direction and scope of the proposed interpretation, that interpretation is retained. When the executed result is correct but features such as skewness, leverage, heterogeneity, or sparsity limit what it represents, the wording is qualified accordingly. If the visible structure conflicts with the proposed interpretation or shows that the statistic is dominated by an unrepresentative pattern, the interpretation is reconsidered and may be rewritten or omitted.

This verification record provides traceability rather than formal statistical validity. A chart may reveal a data or coding anomaly, but its primary role here is not to recompute the statistic; it is to examine whether the statistic adequately represents the relevant data structure. Moreover, the chart and its interpretation are both produced within an agentic pipeline. EDAgent is therefore positioned as an exploratory auditing mechanism, and its findings still require statistical confirmation and expert review when decisions are consequential. Table~\ref{tab:verification_record} specifies what the system retains.

\subsection{System Overview}

\begin{figure}[t]
    \centering
    \includegraphics[width=0.85\linewidth,height=0.42\textheight,keepaspectratio]{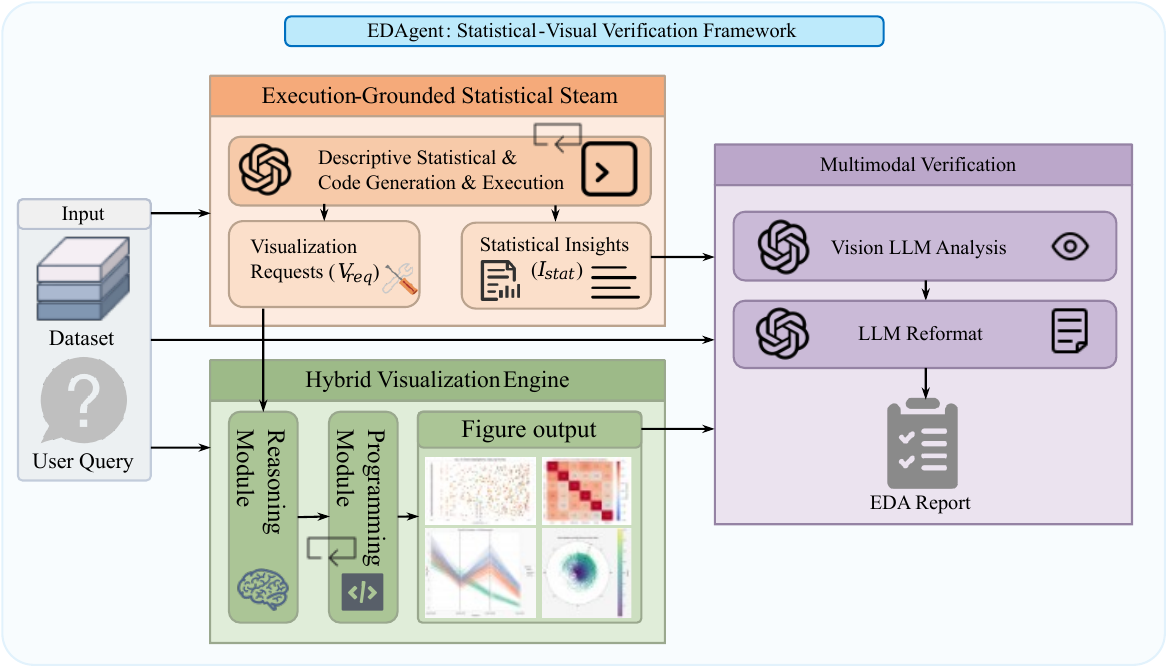}
    \caption{Overview of the evaluated EDAgent implementation. Execution-grounded analysis produces statistical findings and verification requests; the visualization engine generates targeted charts; and multimodal verification determines how the evidence enters the final report.}
    \label{fig:overall}
\end{figure}

Figure~\ref{fig:overall} presents the complete workflow. The statistical stream generates and executes descriptive-analysis code to establish numerical evidence. The visualization engine converts selected findings into rendered charts through general-purpose reasoning, specialized visualization planning, and executable plotting code. The final stage assigns a verification outcome and synthesizes the resulting interpretation into a structured report. The present implementation is sequential: visual findings do not trigger a new round of statistical-code generation or chart regeneration. The particular allocation of GPT-OSS-120B, GPT-5.1, and VisPlan to these stages describes the evaluated implementation; the verification workflow itself does not require these specific models.

\paragraph*{Execution-Grounded Statistical Stream ($\Phi_{\mathrm{stat}}$)}
GPT-OSS-120B generates descriptive-analysis code from the dataset context and query. The code is executed locally, and the resulting standard output forms the numerical evidence $I_{\mathrm{stat}}$. GPT-5.1 then summarizes the executed results and formulates visual verification requests $V_{\mathrm{req}}$ \citep{openai2025gptoss,openai2025gpt51}.

\paragraph*{Hybrid Visualization Engine ($\Phi_{\mathrm{vis}}$)}
GPT-OSS-120B converts the analytical context and $V_{\mathrm{req}}$ into visualization rationales. VisPlan translates these rationales into fine-grained visualization plans, and GPT-5.1 generates and executes plotting code to produce a collection of rendered charts $\mathcal{P}=\{p_1,p_2,\ldots,p_k\}$.

\paragraph*{Multimodal Verification and Report Synthesis ($\Phi_{\mathrm{verify}}$)}
GPT-5.1 compares the rendered figures with the statistical findings and determines whether the initial interpretation should be retained, qualified, or reconsidered. The statistical and visual findings are then synthesized and formatted into the final report $R$.

\subsection{Execution-Grounded Statistical Analysis}

The statistical stream establishes an explicit numerical basis for the subsequent analysis. Rather than asking an LLM to directly produce numerical conclusions from raw tabular context, the system first generates executable descriptive-analysis code and grounds all downstream statistical statements in the resulting program output.

\subsubsection{Code Generation and Local Execution}

GPT-OSS-120B receives the user query $Q$, dataset descriptions, file paths, file-level metadata, inferred data types, and previews of the input tables. It is instructed to generate Python code for descriptive exploratory analysis, including operations such as dataset profiling, missing-value inspection, distribution summaries, grouped comparisons, and association analysis when relevant to the query. The generated program loads the original data files and prints its numerical results to standard output.

The code is executed in a local Python subprocess separate from the language-model response. If execution fails, the error message and the unsuccessful code are returned to GPT-5.1 for revision. This repair-and-execution process continues until the program succeeds or a predefined maximum number of attempts is reached. The successful output is recorded as $I_{\mathrm{stat}}$.

\subsubsection{Statistical Findings and Verification Requests}

GPT-5.1 receives $I_{\mathrm{stat}}$ together with the user query and relevant dataset context. It converts the execution output into a concise set of statistical findings while preserving the values produced by the program. The model is instructed not to introduce numerical quantities that are absent from the execution output.

The same stage generates $V_{\mathrm{req}}$, a set of visual verification requests designed to examine or contextualize the statistical findings. These requests target potential limitations of aggregate summaries, including outlier sensitivity, skewness, multimodality, subgroup heterogeneity, nonlinear relationships, and interactions among variables. The requests do not prescribe all rendering parameters; instead, they identify the statistical findings and data relationships that require visual examination.

\subsection{Hybrid Visualization Engine}

\begin{figure}[t]
    \centering
    \includegraphics[width=0.85\linewidth,height=0.42\textheight,keepaspectratio]{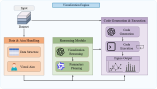}
    \caption{Hybrid visualization pipeline. GPT-OSS-120B generates visualization rationales, VisPlan converts them into structured plans, and GPT-5.1 generates and iteratively repairs executable plotting code.}
    \label{fig:inference_process}
\end{figure}

The hybrid visualization engine converts verification requests into rendered charts through three stages: visualization rationale generation, fine-grained planning with VisPlan, and iterative plotting-code execution. Figure~\ref{fig:inference_process} summarizes this process.

\subsubsection{Visualization Rationale Generation}

GPT-OSS-120B receives the dataset description, user query $Q$, file-level metadata, data-type information, previews of the input tables, and the verification requests $V_{\mathrm{req}}$. Based on this context, it generates a visualization rationale describing the analytical purpose, relevant variables, and proposed chart type for each figure. The model is instructed to include the charts required by $V_{\mathrm{req}}$ and may additionally propose exploratory visualizations that could reveal patterns not captured by the descriptive statistics.

\subsubsection{Fine-Grained Visualization Planning with VisPlan}

The visualization rationales for all requested figures in an analysis instance are submitted to VisPlan together with the relevant dataset and analysis context. VisPlan generates a joint structured plan for the complete figure set, including data mappings, chart types, figure dimensions, labels, legends, color configurations, layout choices, and annotations where appropriate. Generating the figure plans in a single request allows design choices to be coordinated across charts produced for the same EDA report.

The rationale and plan are concatenated to form the specification supplied to the downstream code-generation model. GPT-OSS-120B determines the analytical purpose of the figures, whereas VisPlan provides the fine-grained configuration used to implement them.

\subsubsection{Plotting-Code Generation and Execution}

GPT-5.1 receives the combined visualization rationale and plan, together with the dataset paths, file names, and output requirements. It generates Python code that loads the input tables and saves the rendered charts as image files. The implementation uses a common plotting environment across all generated figures and excludes interactive backends that are unsuitable for the report-generation workflow.

The generated code is executed in a local subprocess. If execution fails, the error message is appended to the code-generation conversation and returned to GPT-5.1 for revision. This repair-and-execution process continues until the script succeeds or the predefined retry limit is reached. If repeated incremental revisions fail, the system can discard the unsuccessful code-generation context and request a new implementation from the original visualization rationale and plan.

\subsection{Multimodal Verification and Report Synthesis}

The final stage compares the rendered visualizations with the findings derived from the execution-grounded statistical analysis. Rather than treating the statistical and visual outputs as independent report components, GPT-5.1 examines whether the visible data patterns warrant retaining, qualifying, or reconsidering the initial interpretation and records the corresponding $F_j$, $O_j$, and $R_j$ fields.

\subsubsection{Multimodal Verification}

After the visualization collection $\mathcal{P}=\{p_1,p_2,\ldots,p_k\}$ has been generated, GPT-5.1 receives the user query $Q$, the rendered charts, and the statistical findings derived from $I_{\mathrm{stat}}$. The model performs three complementary forms of analysis.

\paragraph*{Verification Outcome}
The model compares each visible pattern with the statistical finding that motivated the chart and determines whether the interpretation should be retained, qualified, or reconsidered. The assessment must state which feature of the chart determines the outcome, such as skewness, an influential observation, subgroup variation, sparse cells, or nonlinear structure.

\paragraph*{Additional Exploratory Findings}
The model identifies visually observable patterns that were not represented in the descriptive statistical summary. Such findings may include multimodal distributions, clusters, threshold-like effects, heterogeneous subgroup relationships, or interactions among multiple variables. These observations are treated as exploratory rather than confirmatory or causal conclusions.

\paragraph*{Visual Evidence Grounding}
Each visually derived finding is required to identify the chart from which the evidence was obtained and the statistical result to which it responds. This constraint makes the path from computation to chart to revised statement traceable.

The output of this stage is a finding-level visual assessment $I_{\mathrm{vis}}$ that records the verification outcome and summarizes additional structures identified through visual inspection. This output is an exploratory interpretation of a generated chart; it does not imply an independent validation sample or a calibrated visual test.

\subsubsection{Evidence-Grounded Report Synthesis}

GPT-5.1 combines the statistical findings derived from $I_{\mathrm{stat}}$, the finding-level visual assessment $I_{\mathrm{vis}}$, the original query $Q$, and references to the generated figures. It organizes these inputs into a report containing an executive summary, descriptive statistical findings, visual verification and exploratory evidence, and conclusions with potential directions for subsequent analysis.

A final formatting pass restructures numerical results into readable tables, aligns the executive summary with the analysis objective, embeds the generated figures in the appropriate sections, and standardizes the presentation. This pass is restricted to organization, wording, and formatting and is instructed not to introduce new numerical values or unsupported analytical conclusions.

\subsection{Implementation Component: Training VisPlan}

VisPlan is designed to convert a high-level visualization rationale into a structured plan that can directly guide plotting-code generation. To develop this capability, we construct a training corpus from highly voted public data-analysis notebooks and refine the extracted visualization configurations using rule-based design guidance. The resulting dataset is used for parameter-efficient fine-tuning of Qwen3-8B \citep{yang2025qwen3}.

\subsubsection{Notebook Collection and Filtering}

We selected the 100 public datasets with the highest numbers of community votes on Kaggle. For each dataset, we collected its 50 highest-voted notebooks, yielding an initial corpus of 5,000 notebooks. We then removed 747 notebooks that contained no visualization content. The remaining 4,253 notebooks formed the source corpus for constructing the VisPlan training data.

\subsubsection{Visualization Rationale and Plan Extraction}

For each retained notebook, an LLM reconstructs two complementary components: a visualization rationale and a visualization plan. Notebook files are first converted to scripts while preserving cell boundaries and rendering Markdown cells as comments. The extraction implementation queries the \nolinkurl{mistralai/devstral-2512:free} endpoint through OpenRouter. Its prompt requires separate reasoning and plan blocks covering every chart found in the notebook. The rationale describes the analytical objective, variables, and reason for selecting each chart; the plan records the corresponding chart type, data mappings, dimensions, labels, legends, colors, layout, annotations, and other rendering parameters. The extraction call does not explicitly override sampling parameters, so the endpoint defaults govern this stage.

This process transforms heterogeneous notebook code into a unified intermediate representation. Instead of training VisPlan to reproduce complete plotting programs, the corpus focuses on the planning decisions that connect an analytical objective to a concrete visualization configuration. Library-specific code generation remains the responsibility of the downstream programming model.

\subsubsection{Rule-Guided Plan Refinement and Fine-Tuning}

Visualization parameters extracted from public notebooks may reflect default library settings or inconsistent design conventions. We therefore append a fixed set of heuristic rules, distilled from professional data-visualization tutorials, to an LLM refinement prompt. Given an extracted plan, the model rewrites its visualization parameters to follow the rules while preserving the analytical objective, variables, and figure inventory. The rules address chart-type suitability, figure dimensions, label readability, legend placement, color selection, annotations, and visual clutter. The purpose is not to impose a single visual style, but to reduce recurring low-quality configurations and improve the consistency of the supervisory signals.

The resulting 4,253 examples are converted into a consistent structured format for supervised fine-tuning. Each example provides the relevant analytical context and visualization rationale and uses the refined visualization plan as the target output. The implementation uses the \nolinkurl{Qwen/Qwen3-8B} checkpoint and a 90:10 training--validation split generated with split seed 42. VisPlan is trained as a LoRA adapter with rank 16, scaling factor 32, and dropout 0.05. Adapters are applied to the query, key, value, and output projections and to the gate, up, and down projections. Sequences are padded or truncated to 8,192 tokens; prompt and padding tokens are masked from the loss. Training runs for six epochs with per-device batch size 1, gradient accumulation over eight steps, learning rate $2\times10^{-4}$, bfloat16 arithmetic, and gradient checkpointing. The distributed configuration uses full-shard FSDP with automatic transformer-layer wrapping, and evaluation and checkpointing occur every 50 optimization steps. During inference, VisPlan receives the rationale produced by GPT-OSS-120B and generates the structured plan used by GPT-5.1 for plotting-code generation.

\section{Evaluation}

The evaluation has two distinct aims. EDA-Bench describes how often complete EDAgent reports satisfy the proposed evidence rubric under a fixed multimodal judge; because no end-to-end baseline is evaluated, this part is an absolute system audit rather than a comparative experiment. The adapted VisEval study compares visualization planners while holding the downstream code-generation and rendering pipeline fixed. The reported scores and mean differences are descriptive: the protocol uses one judge call per output and provides no repeated-run or sampling-uncertainty analysis. Case studies then show what the aggregate rubric scores cannot: how a visual check changes the wording and scope of a specific statistical finding. The evaluation therefore combines broad rubric-based summaries with case-level evidence.

\subsection{End-to-End Evaluation on EDA-Bench}

\subsubsection{Benchmark Construction}

To evaluate EDAgent on datasets not used to construct the VisPlan training corpus, we created EDA-Bench from public datasets on Kaggle. The collection script orders datasets by community votes and retains only entries whose listed files all use the CSV format. In the ranking snapshot used for this study, the VisPlan corpus was derived from the first 100 eligible datasets, whereas the collection script enumerated the first 200 and retained eligible ranks 101--200 as candidates for EDA-Bench. This dataset-level separation reduces direct overlap between visualization-plan training and end-to-end evaluation; it does not rule out overlap in domains, schemas, plotting conventions, or knowledge acquired during pretraining.

We screened the 100 candidate datasets and excluded 13 that exhibited severe corruption, insufficient usable content, or substantial overlap with datasets included in the VisPlan corpus. The resulting benchmark contains 87 public tabular datasets from Kaggle. It includes applied datasets from domains such as finance, healthcare, and e-commerce, together with canonical instructional datasets such as Iris. The datasets exhibit heterogeneous schemas, variable types, sizes, and distributional characteristics.

For each dataset, EDAgent receives the associated dataset description, an analytical query, and one or more tabular data files. The system executes the complete pipeline without manual intervention in the intermediate statistical analysis, visualization planning, chart generation, multimodal verification, or report-synthesis stages. The resulting report and visualizations are evaluated jointly.

\subsubsection{Multimodal Evaluation Protocol and Metrics}

We evaluate each EDA-Bench output using GPT-5.1 as a multimodal judge. For each dataset, the judge receives sampled computational metadata derived from the underlying tables, the final EDA report generated by EDAgent, and all rendered visualizations referenced in the report. Each report is evaluated once using a fixed prompt and a six-dimensional rubric. Every dimension is scored on an ordinal scale from 1 to 5, where higher scores indicate stronger performance. The judge returns the scores and corresponding explanations in a structured format. GPT-5.1 also participates in EDAgent's plotting and verification stages, so this evaluation is not independent of the system and may reward shared model preferences. We use it as a scalable diagnostic audit, not as a human gold standard.

\paragraph*{Data Factuality and Visual Consistency}
This dimension evaluates whether numerical statements in the report are consistent with the sampled dataset metadata and whether the textual interpretations agree with the rendered charts. It penalizes invented values, incorrect numerical comparisons, and inconsistencies between the report and its visual evidence.

\paragraph*{Critical Data Caveats}
This dimension measures whether the report identifies data-quality or interpretation risks that could materially affect the analysis, including missing values, extreme observations, class imbalance, limited sample sizes, unusual distributions, or other relevant limitations.

\paragraph*{Statistical--Visual Cross-Validation}
This dimension evaluates whether the report uses visual evidence to verify or contextualize statistical findings. A high score requires more than presenting charts alongside numerical results; the report must explicitly explain how a visible pattern affects the interpretation of the corresponding statistical finding.

\paragraph*{Analytical Narrative}
This dimension assesses whether the report develops a coherent analytical progression from dataset characteristics and descriptive statistics to visual exploration, integrated interpretation, and conclusions. It also considers whether the findings are organized around the user query rather than presented as disconnected observations.

\paragraph*{Counterfactual Visual Uplift}
This dimension asks the judge to estimate the analytical contribution of the generated visualizations by considering a counterfactual report in which the figures are removed. A high score indicates that the charts appear to reveal or clarify patterns---such as clusters, skewness, nonlinear relationships, or subgroup heterogeneity---that would be difficult to infer from the textual statistics alone. Because the no-figure report is not separately generated or evaluated, this score is a model judgment about a counterfactual, not a measured treatment effect.

\paragraph*{Next-Step Hypothesis}
This dimension evaluates whether the report proposes specific and actionable directions for subsequent analysis, such as confirmatory statistical tests, targeted data-quality checks, feature construction, subgroup analysis, or predictive modeling. These suggestions should follow from the evidence presented in the report rather than introducing unrelated speculation.

For each dimension, we report the complete score distribution and the proportion of cases assigned the maximum score of 5. EDA-Bench is used to characterize the absolute performance of EDAgent across the six evaluation dimensions.

\subsubsection{Results on EDA-Bench}

\begin{figure}[tbhp]
    \centering
    \includegraphics[width=\linewidth]{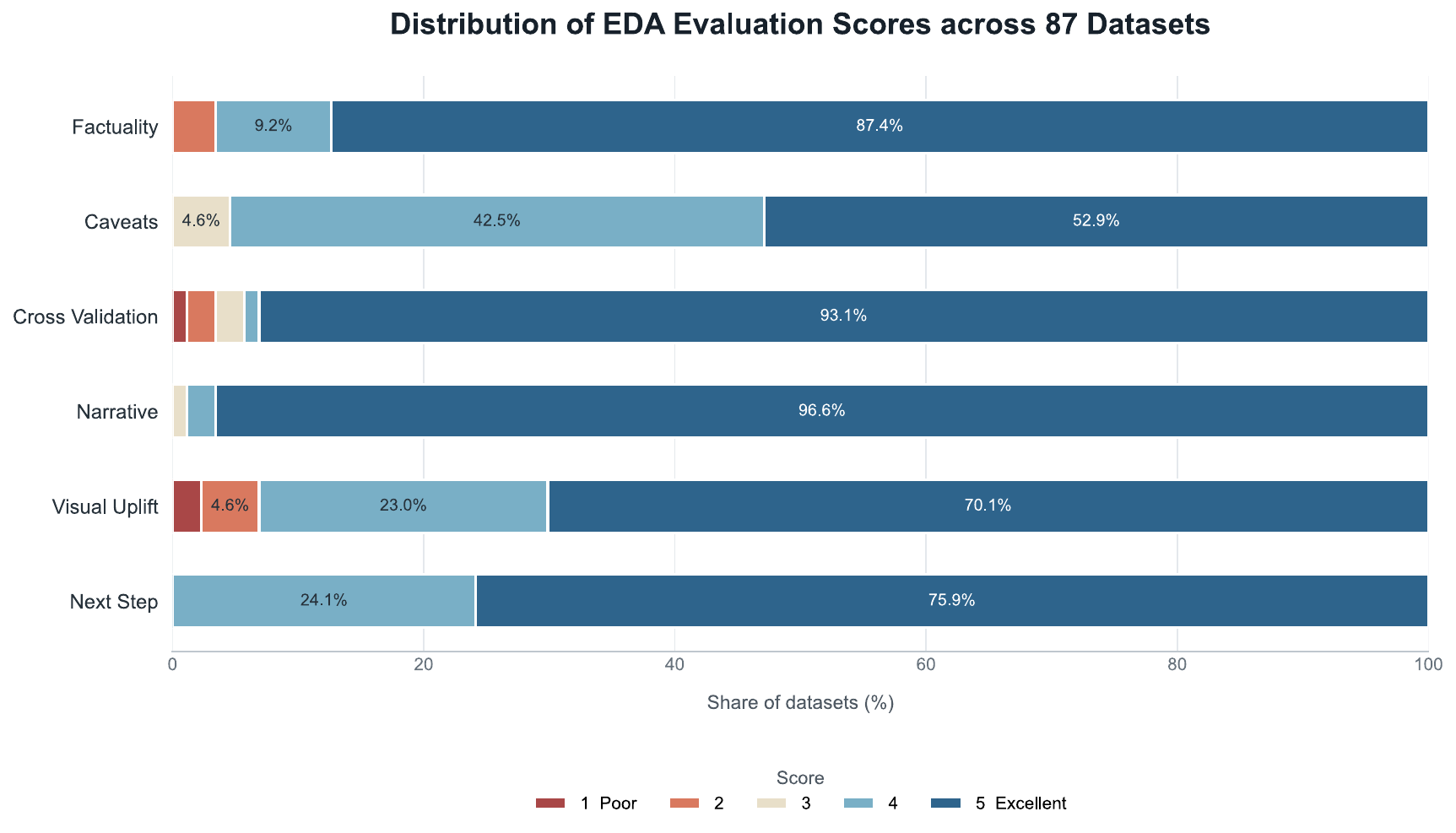}
    \caption{Distribution of GPT-5.1 judge scores across the six EDA-Bench evaluation dimensions for 87 datasets. Each report was evaluated once on a scale from 1 to 5.}
    \label{fig:score_distribution}
\end{figure}

Figure~\ref{fig:score_distribution} presents the score distributions assigned by the GPT-5.1 multimodal judge across the six evaluation dimensions. Overall, the distributions are concentrated toward the upper end of the five-point scale, although the degree of concentration varies across dimensions.

Figure~\ref{fig:score_distribution} shows the complete score distributions, while the following analysis uses a uniform reporting criterion: the proportion of reports assigned the maximum score of 5. The maximum-score rates were 96.6\% for Analytical Narrative, 93.1\% for Statistical--Visual Cross-Validation, 87.4\% for Data Factuality and Visual Consistency, 75.9\% for Next-Step Hypothesis, 70.1\% for Counterfactual Visual Uplift, and 52.9\% for Critical Data Caveats.

Under the adopted rubric, the score distributions were strongest for Analytical Narrative and Statistical--Visual Cross-Validation. Data Factuality and Visual Consistency also received the maximum score for most reports. Because the judge compared each report with sampled computational metadata rather than exhaustively recomputing every reported value, this result reflects agreement with the evidence supplied to the evaluator.

Critical Data Caveats had the lowest maximum-score rate, with 52.9\% of reports receiving a 5. This distribution indicates that reports did not consistently receive full credit for identifying consequential issues such as missingness, extreme observations, imbalance, small subgroups, or unusual distributions. Counterfactual Visual Uplift received a 5 for 70.1\% of reports, indicating that the judge did not always regard every chart as adding substantial information beyond the numerical summary. Together with the higher rates for narrative organization and statistical--visual linkage, these results show that performance varied across evaluation dimensions rather than concentrating uniformly at the ceiling.

Taken together, EDA-Bench shows that the proposed verification structure can be applied across heterogeneous tabular datasets under the adopted rubric.

\subsection{Supporting Component Evaluation on an Adapted VisEval}

\subsubsection{Experimental Setup and Query Adaptation}

We conduct a component-level evaluation of visualization planning using 166 tabular datasets from VisEval \citep{chen2024viseval}. The original benchmark associates tabular datasets with explicit natural-language visualization instructions. For the adapted evaluation used in this study, we construct one open-ended visualization task for each of the 166 datasets, yielding 166 adapted tasks in total.

The original instructions are predominantly imperative and often specify the intended chart or plotting operation directly. Such tasks evaluate whether a model can translate a predefined plotting intent into a corresponding visualization, but they provide limited evidence about whether the model can independently determine an appropriate visual representation for a broader analytical objective.

To evaluate visualization planning in an exploratory setting, we generate one open-ended query for each of the 166 datasets. The new queries are generated by an LLM from the corresponding dataset context and express an analytical objective without prescribing a chart type, visual encoding, attribute mapping, or rendering configuration. For example, an adapted query may ask how participation varies across activities when comparing students and faculty, rather than requesting a specific pie chart. This adaptation changes the construct measured by the original benchmark and may inherit preferences from the query-generation model; results should therefore be interpreted for these 166 adapted tasks, not as a replacement score on the original VisEval benchmark.  

The adapted tasks require the evaluated model to determine which variables should be visualized, which relationships are relevant to the analytical objective, and which chart configuration is suitable for communicating the resulting pattern. The same set of 166 adapted queries is used across all evaluated models.

\subsubsection{Multimodal Evaluation Protocol and Metrics}

For each adapted VisEval task, every evaluated model receives the same dataset context and open-ended analytical query. To focus the comparison on visualization planning, candidate models use a common input format and the same downstream plotting-code generation and execution procedure. The resulting rendered visualization is assessed against the corresponding adapted query.

GPT-5.1 serves as the multimodal judge. For each task, the judge receives the adapted query and the rendered visualization and evaluates the output once using a fixed prompt. It returns scores and brief justifications in a structured format. Each visualization is assessed on four dimensions using an ordinal scale from 1 to 5. As in EDA-Bench, the single-judge design measures one model's assessment and does not quantify judge variability or human agreement.

\paragraph*{Alignment}
Alignment measures whether the visualization addresses the analytical objective expressed by the adapted query. It considers whether the selected variables, chart content, title, axes, legends, and other semantic elements are consistent with the intended analysis.

\paragraph*{Readability}
Readability assesses whether the information encoded in the chart can be extracted without unnecessary difficulty. It considers text legibility, overlapping elements, visual clutter, axis presentation, legend placement, and whether contrast and layout support clear interpretation.

\paragraph*{Effectiveness}
Effectiveness evaluates whether the selected chart type and visual encodings are appropriate for the analytical task. It considers whether position, length, color, size, grouping, and other encodings communicate the relevant comparison, distribution, relationship, or trend efficiently and without avoidable ambiguity.

\paragraph*{Aesthetics}
Aesthetics assesses the overall visual design of the chart, including layout balance, spacing, typographic consistency, color coordination, and the restrained use of decorative elements.

The judge additionally produces an Overall Score summarizing the visualization across the four dimensions. For each model, we report the mean Alignment, Readability, Effectiveness, Aesthetics, and Overall scores across the 166 tasks.

\subsubsection{Baselines and Results}

\begin{figure}[t]
    \centering
    \includegraphics[width=\linewidth]{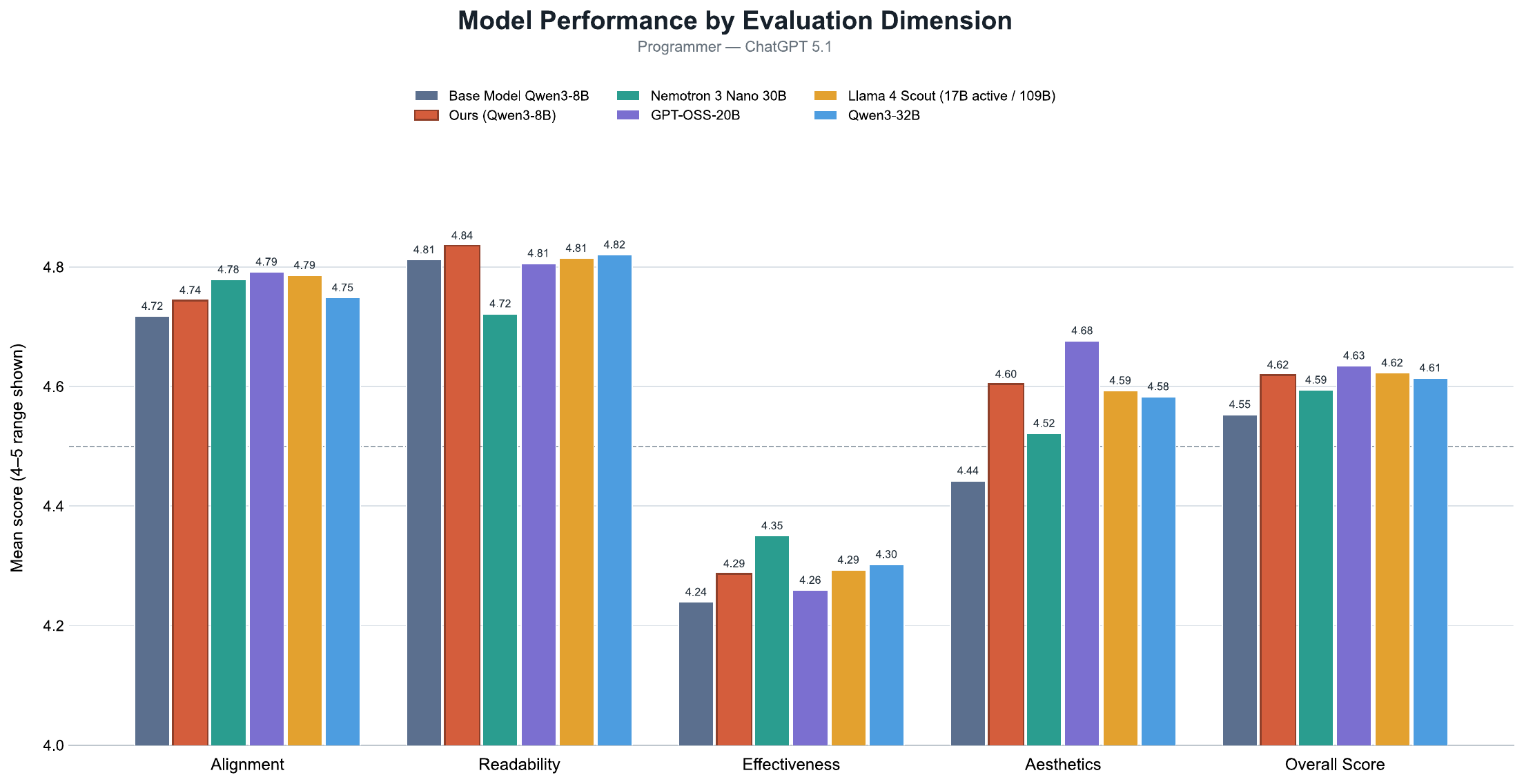}
    \caption{Mean visualization-quality scores on the 166 adapted VisEval tasks. All models use the same open-ended queries, downstream rendering pipeline, and GPT-5.1 evaluation protocol.}
    \label{fig:visualization_evaluation}
\end{figure}

We compare VisPlan, fine-tuned from Qwen3-8B, with its original Qwen3-8B base model and four general-purpose language models: Nemotron 3 Nano 30B, GPT-OSS-20B, Qwen3-32B, and Llama 4 Scout, a mixture-of-experts model with 109B total parameters and approximately 17B active parameters per inference step \citep{nvidia2025nemotron3,openai2025gptoss,yang2025qwen3,meta2025llama4}. All models are evaluated on the same 166 adapted VisEval tasks using an identical downstream code-generation, rendering, and multimodal evaluation pipeline. Figure~\ref{fig:visualization_evaluation} reports the mean scores for the four dimensions and the Overall Score.

VisPlan's observed mean exceeded that of its Qwen3-8B base model on all four dimensions: Alignment increased from 4.72 to 4.74, Readability from 4.81 to 4.84, Effectiveness from 4.24 to 4.29, and Aesthetics from 4.44 to 4.60. The Overall Score increased from 4.55 to 4.62. The gain was concentrated in Aesthetics ($+0.16$); the other changes ranged from $+0.02$ to $+0.05$. This pattern is consistent with the training objective, which refines chart configuration and presentation, but the single-judgment design does not support an inferential claim that every difference would reproduce.

The larger general-purpose baselines had tightly clustered observed Overall Scores: 4.59 for Nemotron 3 Nano 30B, 4.61 for Qwen3-32B, 4.62 for Llama 4 Scout, and 4.63 for GPT-OSS-20B. Llama 4 Scout uses a mixture-of-experts architecture with 109B total parameters and approximately 17B active parameters per inference step. VisPlan's mean of 4.62 lies within this descriptive range. We interpret this as evidence that the compact specialized planner remained competitive under the shared pipeline, not as statistical equivalence to any individual baseline.

\subsection{Case Studies}

To complement the aggregate evaluation, we examine illustrative outputs from the end-to-end and component-level evaluations. The cases are organized around an explicit sequence---statistical finding, targeted visual check, and consequence for the report---so that the role of the plot is visible rather than implied.

\subsubsection{Statistical--Visual Verification Cases}

The first two cases show qualification: the computed summary remains correct, but the chart narrows what can responsibly be said. The third shows retention: the visual structure reinforces the statistical ranking of candidate features. Together, they demonstrate that visual verification can add value without requiring the chart to contradict the executed result.

\paragraph*{Case 1: Outlier-Sensitive Correlation and a Threshold-Like Pattern in Car Sales}

For the passenger-vehicle subset of the car sales dataset, the statistical stream reported a Pearson correlation of approximately $-0.340$ between \textit{Horsepower} and \textit{Sales}. Considered in isolation, this coefficient suggests a moderate negative association between the two variables. The corresponding scatter plot, however, shows that most passenger vehicles occupy a comparatively compact horsepower range, whereas the Dodge Viper appears as an isolated observation with approximately 450 horsepower and comparatively low sales. The Dodge Viper belongs to the same passenger-vehicle subset and is highlighted separately in the legend because its position suggests that it may exert substantial influence on the aggregate association.

\begin{figure}[t]
    \centering
    \begin{minipage}[t]{0.48\linewidth}
        \centering
        \includegraphics[width=\linewidth,height=0.31\textheight,keepaspectratio]{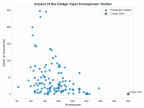}
        \captionsetup{font=small}
        \captionof{figure}{Scatter plot of horsepower and sales within the passenger-vehicle subset. The Dodge Viper, which belongs to this subset, is highlighted separately as an isolated high-horsepower, low-sales observation that is visually positioned to have potentially high leverage on the aggregate correlation.}
        \label{fig:car_sales_outlier}
    \end{minipage}
    \hfill
    \begin{minipage}[t]{0.48\linewidth}
        \centering
        \includegraphics[width=\linewidth,height=0.31\textheight,keepaspectratio]{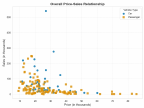}
        \captionsetup{font=small}
        \captionof{figure}{Scatter plot of vehicle price and sales, showing broad sales variation below approximately \$35,000 and a concentration of lower-sales observations above approximately \$40,000.}
        \label{fig:car_sales_threshold}
    \end{minipage}
\end{figure}

By relating the computed correlation to the visual distribution, EDAgent produced a more qualified interpretation. The isolated position of the Dodge Viper suggests that the observed negative association may be sensitive to this observation, while the remaining passenger vehicles form a diffuse point cloud without a comparably clear linear pattern. The visualization therefore does not contradict the computed coefficient; instead, it identifies a potentially influential point that motivates a follow-up sensitivity analysis, such as recomputing the association after excluding the observation or using a robust correlation measure. This is a summary--structure mismatch: a valid aggregate statistic supports an overly broad interpretation until the underlying distribution is examined.

For a separate analysis of vehicle price and sales, EDAgent used the full vehicle dataset rather than the passenger-only subset. The plotted observations therefore include both the \textit{Car} and \textit{Passenger} categories shown in the figure legend. The statistical stream reported an overall Pearson correlation of approximately $-0.30$, indicating a weak-to-moderate negative linear association. The global scatter plot revealed a more structured pattern: below approximately \$35,000, sales varied widely and included several high-volume models, whereas observations above approximately \$40,000 were concentrated in the lower sales range.

This threshold-like pattern is not fully represented by a single linear correlation coefficient. Together, the two analyses demonstrate how visual verification adds distributional context to statistical summaries: the passenger-only horsepower--sales chart exposes sensitivity to an influential observation, while the full-dataset price--sales chart reveals a nonuniform relationship across different price ranges.

\paragraph*{Case 2: Mean Differences, Distributional Skew, and Joint Subgroup Patterns in Housing Prices}

In the housing prices dataset, the statistical stream found that the mean house price increased across several bedroom-count groups. In particular, the mean price increased from approximately 4.95 million for three-bedroom properties to 5.73 million for four-bedroom properties in the dataset's reported units. Considered alone, these group means could suggest that an additional bedroom is associated with a broad upward shift in property values.

\begin{figure}[t]
    \centering
    \begin{minipage}[t]{0.48\linewidth}
        \centering
        \includegraphics[width=\linewidth,height=0.31\textheight,keepaspectratio]{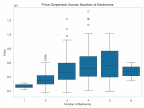}
        \captionsetup{font=small}
        \captionof{figure}{House-price distributions grouped by bedroom count. The higher mean prices in several groups are accompanied by increased dispersion and high-value observations, whereas the shifts in the medians are comparatively modest.}
        \label{fig:housing_boxplot}
    \end{minipage}
    \hfill
    \begin{minipage}[t]{0.48\linewidth}
        \centering
        \includegraphics[width=\linewidth,height=0.31\textheight,keepaspectratio]{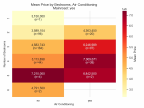}
        \captionsetup{font=small}
        \captionof{figure}{Mean house prices by bedroom count and air-conditioning availability for properties with main-road access (\texttt{mainroad=yes}). Cell means should be interpreted together with their sample sizes because several bedroom--air-conditioning combinations contain very few observations.}
        \label{fig:housing_heatmap}
    \end{minipage}
\end{figure}

The corresponding boxplots provide a more detailed view of the group distributions. As shown in Figure~\ref{fig:housing_boxplot}, the higher means for the three- and four-bedroom groups are accompanied by wider price ranges and several high-value observations. In contrast, the changes in the group medians are more modest than the changes in the means.

The visual evidence therefore qualifies the interpretation based on group means. The observed increase does not correspond to a uniform upward shift of the entire price distribution. Instead, part of the difference is associated with a relatively small number of expensive properties concentrated in the three- and four-bedroom groups. The number of bedrooms is related to price in this dataset, but the magnitude of the mean difference is sensitive to the upper tail of the distribution. This is a second summary--structure mismatch: the means overstate the extent to which the full distributions have shifted.

EDAgent further examined bedroom count and air-conditioning availability within the subset of properties that have main-road access. Figure~\ref{fig:housing_heatmap} shows that the largest displayed mean occurs in the five-bedroom, no-air-conditioning cell (7.21 million), but this estimate is based on only four observations. Among groups with moderate or large sample sizes, the air-conditioned three-bedroom and four-bedroom groups have relatively high mean prices, approximately 6.25 million ($n=97$) and 7.01 million ($n=38$), respectively. Several edge cells are sparse---including one-bedroom properties without air conditioning ($n=1$), five-bedroom properties with air conditioning ($n=2$), five-bedroom properties without air conditioning ($n=4$), and six-bedroom properties without air conditioning ($n=2$)---so their group means should be interpreted cautiously.

Within the \texttt{mainroad=yes} subset, the heatmap reveals joint subgroup patterns but does not by itself establish a statistical interaction between bedroom count and air-conditioning availability. The observed differences suggest possible effect modification or subgroup heterogeneity that should be tested using a model containing an interaction term, for example \textit{price} $\sim$ \textit{bedrooms} $+$ \textit{airconditioning} $+$ \textit{bedrooms} $\times$ \textit{airconditioning}, together with relevant covariates. A follow-up analysis should also compare the same subgroups among properties without main-road access and account for the instability of estimates from sparse cells.

\paragraph*{Case 3: Feature Separation and Candidate Predictors in the Iris Dataset}

The Iris dataset, included in EDA-Bench as a canonical instructional dataset, provides a complementary example in which the statistical and visual evidence are broadly consistent. The statistical stream identified \textit{petal\_length} and \textit{petal\_width} as the variables with the largest differences in mean values across the three species, with maximum between-species mean gaps of approximately 4.088 and 1.782, respectively. The corresponding differences for the sepal measurements were smaller.

\begin{figure}[t]
    \centering
    \begin{minipage}[t]{0.48\linewidth}
        \centering
        \includegraphics[width=\linewidth,height=0.31\textheight,keepaspectratio]{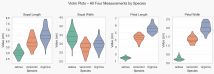}
        \captionsetup{font=small}
        \captionof{figure}{Distributions of the four Iris measurements across species. Petal length and petal width provide substantially stronger between-species separation than the two sepal measurements.}
        \label{fig:iris_violin}
    \end{minipage}
    \hfill
    \begin{minipage}[t]{0.48\linewidth}
        \centering
        \includegraphics[width=\linewidth,height=0.31\textheight,keepaspectratio]{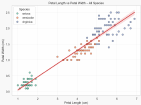}
        \captionsetup{font=small}
        \captionof{figure}{Scatter plot of petal length and petal width across the three Iris species. The two petal measurements form compact species-related clusters with substantially less overlap than the sepal measurements.}
        \label{fig:iris_scatter}
    \end{minipage}
\end{figure}

The violin plots in Figure~\ref{fig:iris_violin} provide the distributional context underlying these mean differences. Petal length and petal width exhibit substantially greater separation among the three species, with the \textit{setosa} distribution clearly distinguished from the other two species. By comparison, the sepal-length and sepal-width distributions show considerably greater overlap.

A two-dimensional scatter plot of petal length and petal width further illustrates their joint discriminative structure. As shown in Figure~\ref{fig:iris_scatter}, the observations form three compact species-related clusters. The \textit{setosa} samples are clearly separated, while the \textit{versicolor} and \textit{virginica} clusters exhibit limited overlap.

The combined statistical and visual evidence identifies the petal measurements as the variables showing the largest raw between-species mean differences and the clearest visual separation. Based on this structure, EDAgent recommends beginning subsequent modeling with interpretable linear classification baselines, such as multinomial logistic regression or a linear support vector machine, using the petal variables as primary features.

Detailed reports of the case studies are available at the anonymous link:
\url{https://anonymous.4open.science/r/EDAgent-Report-551C/}.

\subsubsection{Visualization Planning Cases}

To complement the aggregate adapted VisEval results, we present two illustrative comparisons of visualizations produced using VisPlan, the original Qwen3-8B model, and Qwen3-32B as the planning component. The downstream code-generation and rendering procedure is unchanged across the three models. These selected cases explain the kind of difference captured by the aesthetics score; they are not a substitute for the aggregate evaluation and should not be read as a frequency estimate.

\begin{figure*}[t]
    \centering

    \begin{subfigure}[b]{0.32\textwidth}
        \centering
        \includegraphics[width=\linewidth]{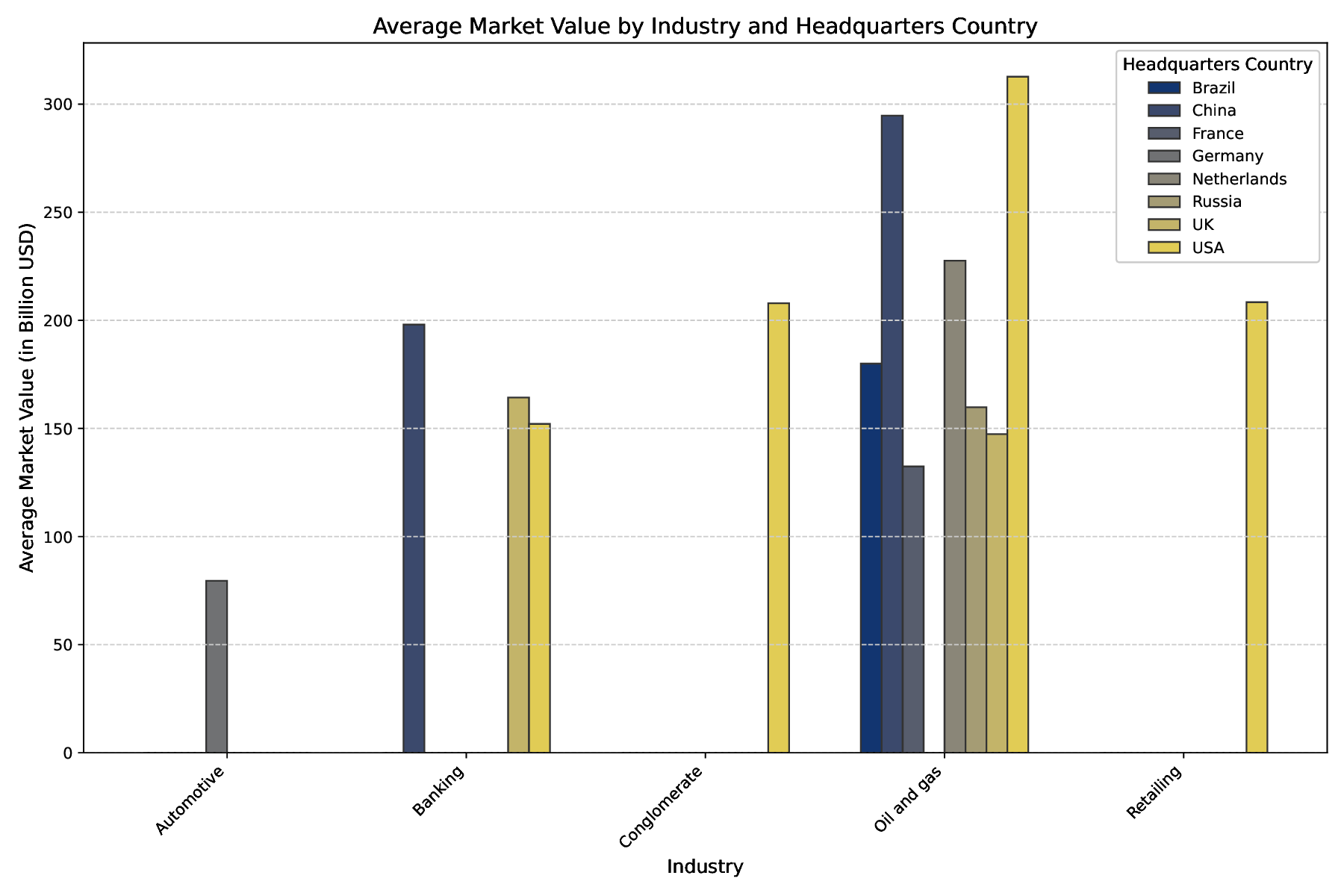}
        \caption{VisPlan}
        \label{fig:vis_case1_visplan}
    \end{subfigure}
    \hfill
    \begin{subfigure}[b]{0.32\textwidth}
        \centering
        \includegraphics[width=\linewidth]{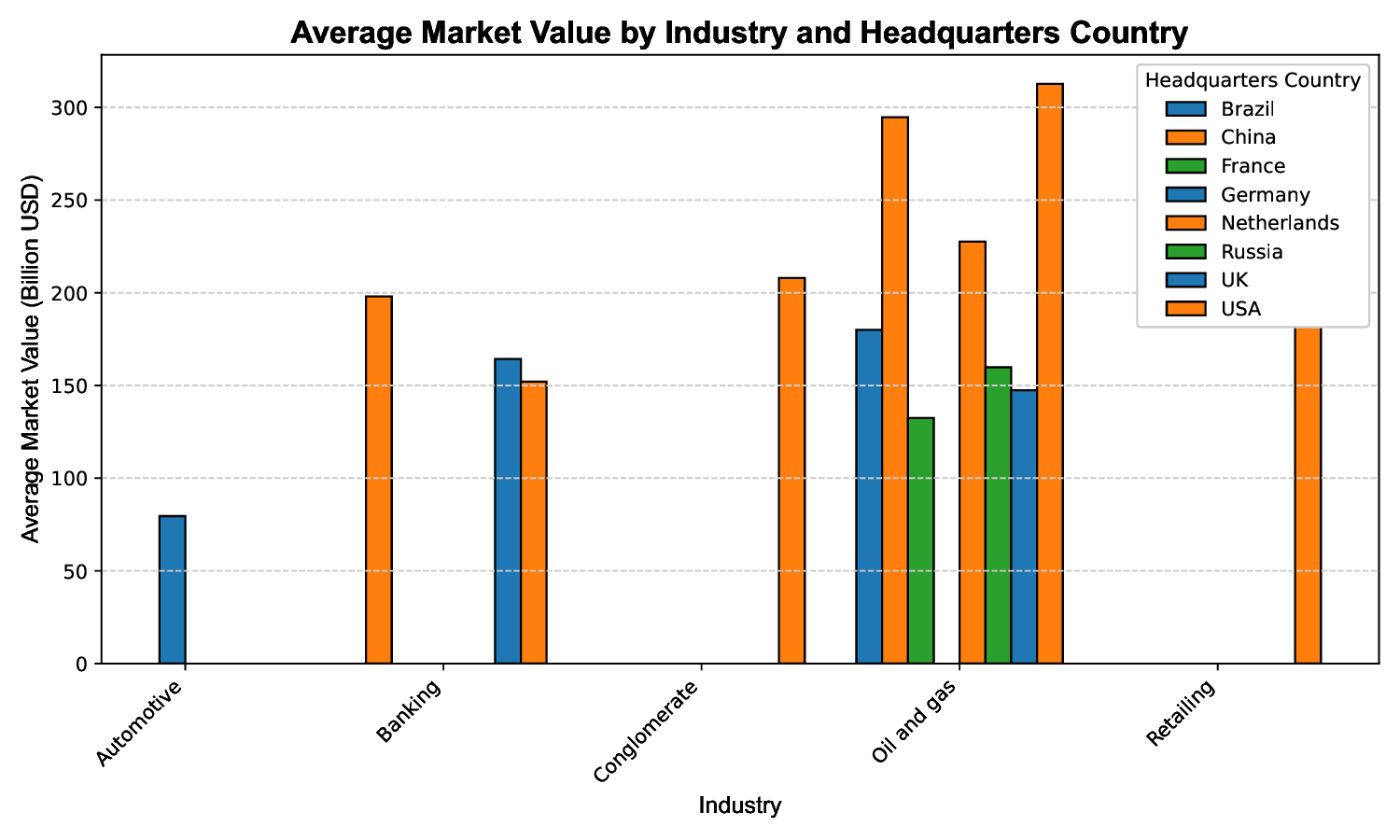}
        \caption{Qwen3-8B}
        \label{fig:vis_case1_8b}
    \end{subfigure}
    \hfill
    \begin{subfigure}[b]{0.32\textwidth}
        \centering
        \includegraphics[width=\linewidth]{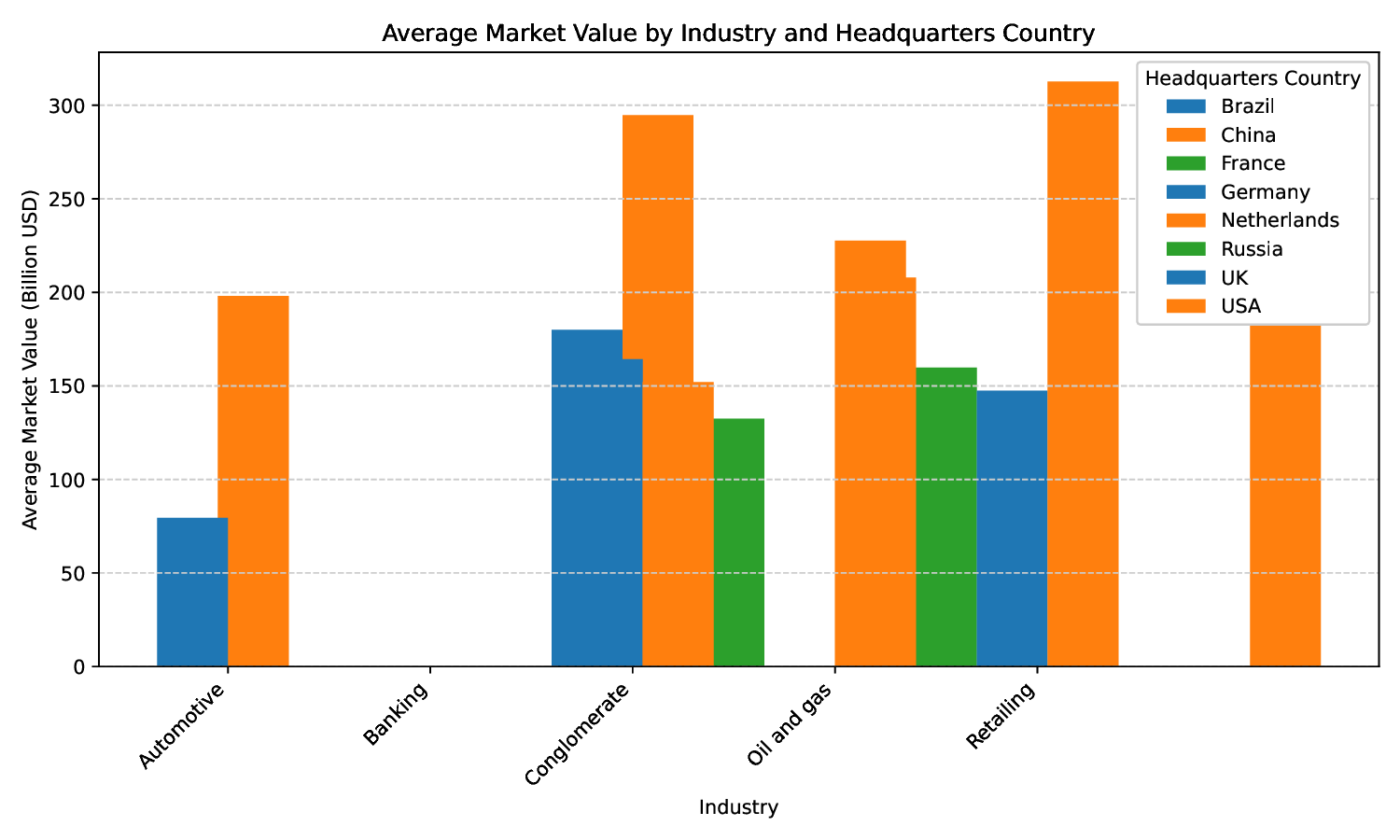}
        \caption{Qwen3-32B}
        \label{fig:vis_case1_32b}
    \end{subfigure}

    \caption{Comparison of visualizations for average market value by industry and headquarters. VisPlan applies a subdued color palette and explicit styling parameters, whereas the two general-purpose models rely more heavily on default categorical colors.}
    \label{fig:vis_case_1}
\end{figure*}

Figure~\ref{fig:vis_case_1} illustrates differences in the parameterization of an individual chart. The visualization generated from the VisPlan specification uses explicitly selected low-saturation colors together with adjusted grid transparency, spacing, and legend styling. By contrast, the Qwen3-8B and Qwen3-32B outputs primarily use the default categorical color cycles provided by the plotting library. The differences are concentrated not in the underlying data mappings, which are similar across the three outputs, but in the degree of control over the final visual presentation.

\begin{figure*}[t]
    \centering

    \begin{subfigure}[b]{0.32\textwidth}
        \centering
        \includegraphics[width=\linewidth]{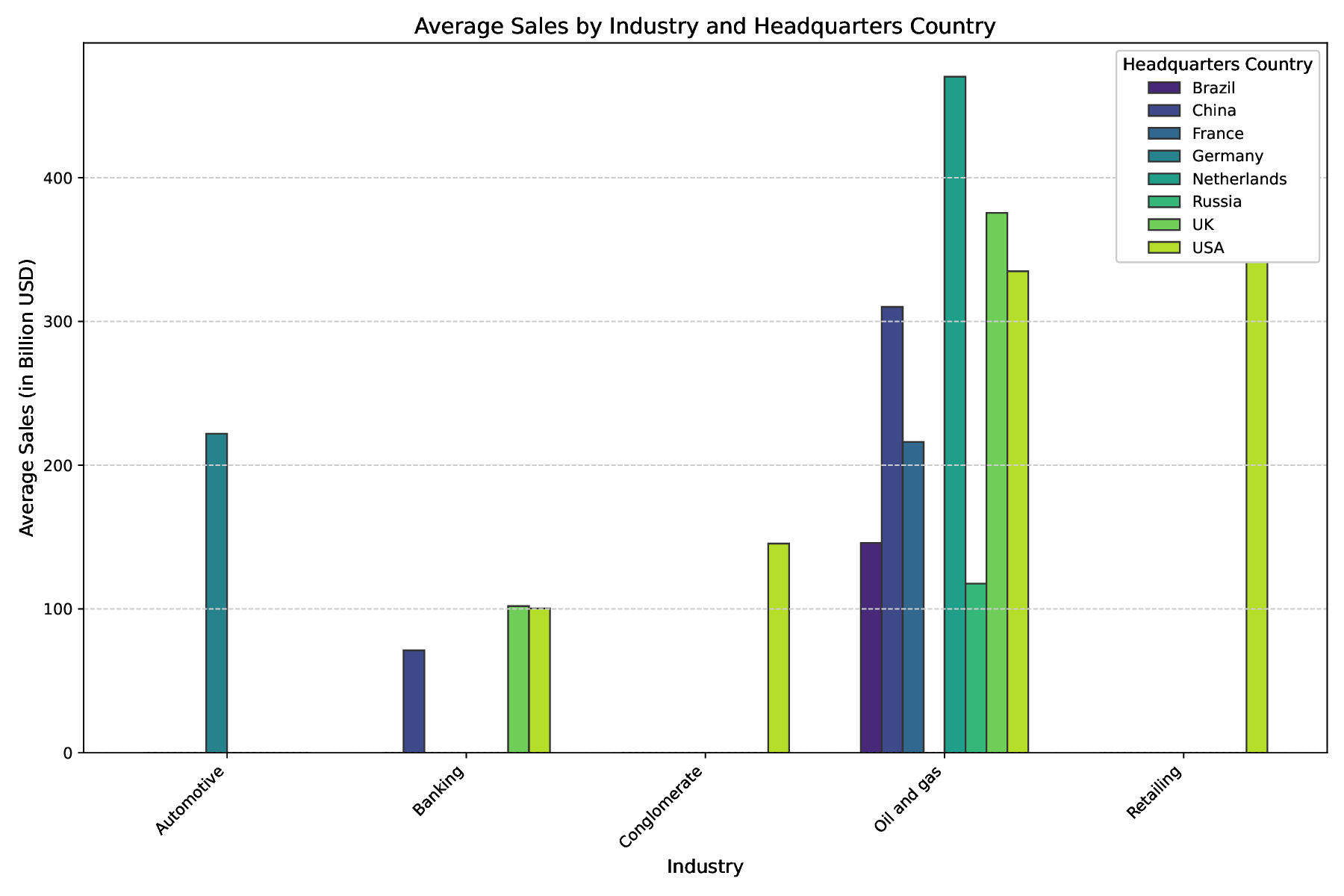}
        \caption{VisPlan}
        \label{fig:vis_case2_visplan}
    \end{subfigure}
    \hfill
    \begin{subfigure}[b]{0.32\textwidth}
        \centering
        \includegraphics[width=\linewidth]{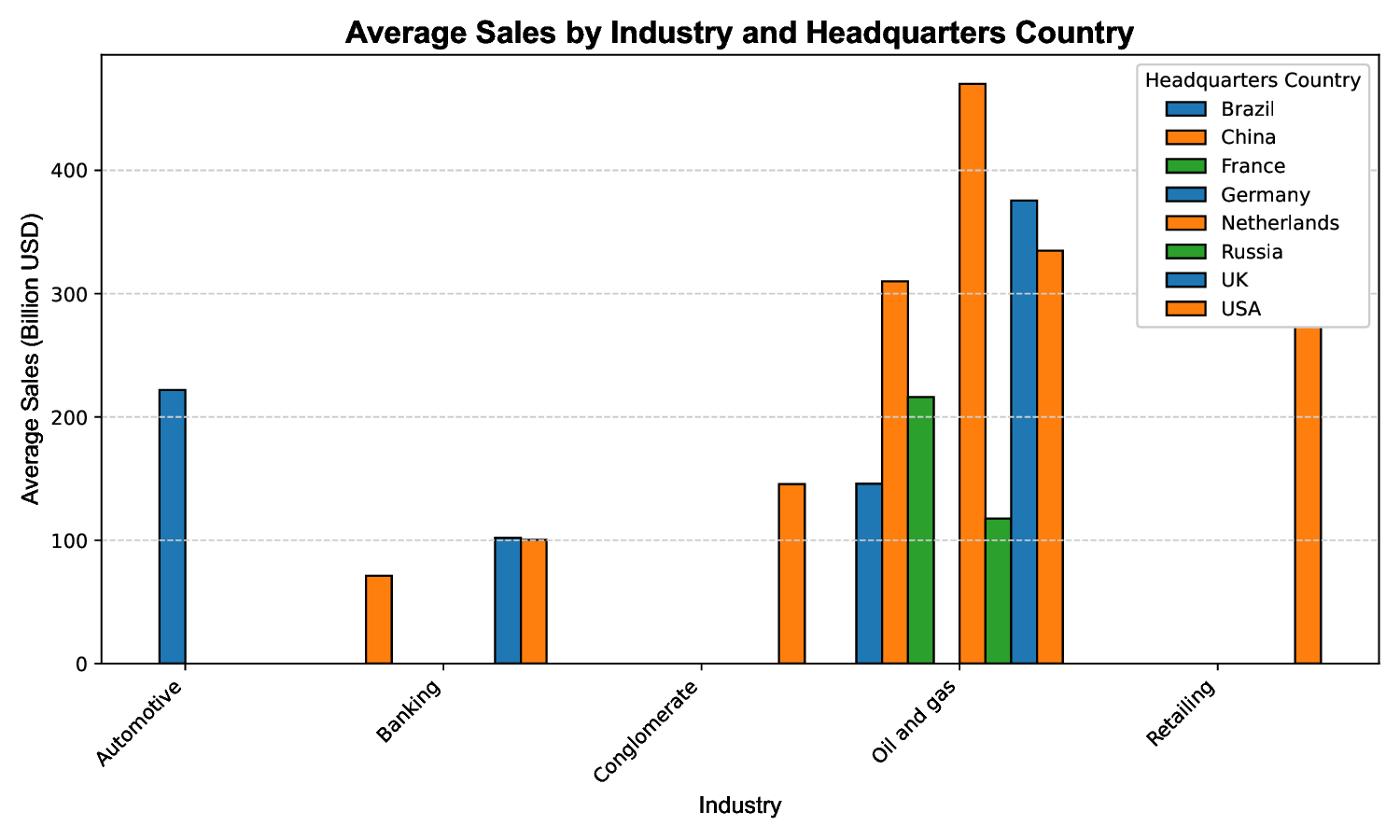}
        \caption{Qwen3-8B}
        \label{fig:vis_case2_8b}
    \end{subfigure}
    \hfill
    \begin{subfigure}[b]{0.32\textwidth}
        \centering
        \includegraphics[width=\linewidth]{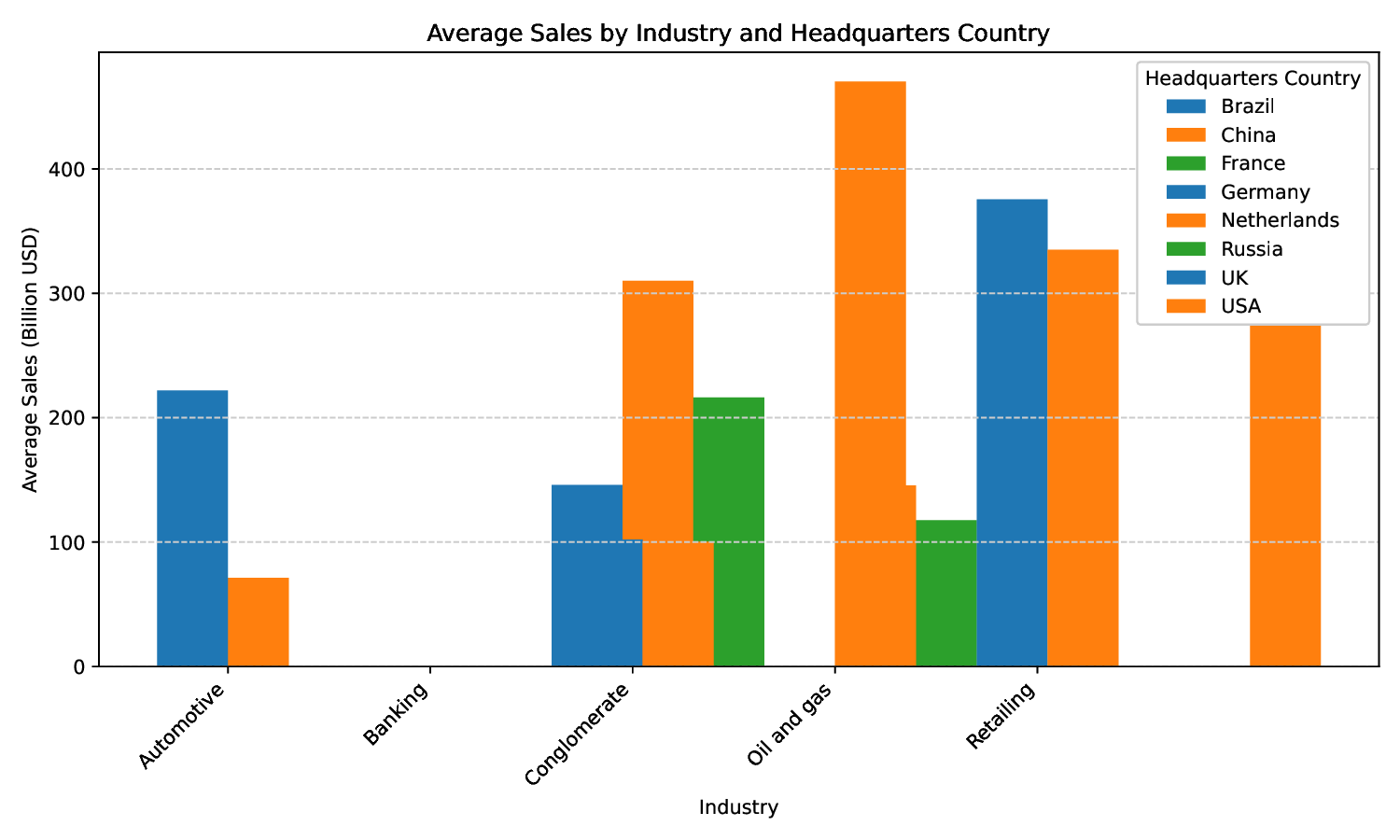}
        \caption{Qwen3-32B}
        \label{fig:vis_case2_32b}
    \end{subfigure}

    \caption{Comparison of visualizations for average sales by industry and headquarters. VisPlan uses a palette distinct from that of the market-value chart, while the two general-purpose models reuse similar default color cycles across the two tasks.}
    \label{fig:vis_case_2}
\end{figure*}

The second comparison highlights parameter coordination across related visualizations. For the average-sales task, VisPlan changes the color palette while retaining comparable layout and labeling conventions. This distinguishes the sales visualization from the preceding market-value visualization without disrupting the overall presentation. In contrast, both Qwen3-8B and Qwen3-32B reuse similar blue--orange--green color cycles for the two metrics.

Together, these examples show that VisPlan affects both local chart parameters and the coordination of design choices across multiple figures. The qualitative differences are consistent with the larger improvement observed in the Aesthetics dimension of adapted VisEval.

\section{Discussion}

\subsection{Visualization as a Verification Step}

The central contribution of EDAgent is a change in the role assigned to a generated chart. In a conventional report pipeline, computation produces findings and visualization illustrates them. In EDAgent, a finding also produces a verification request that specifies which visible feature could confirm its scope or reveal that it is sensitive to other aspects of the distribution. The car-sales and housing cases show the practical effect. The statistical result is retained, but the report changes from a broad description to a conditional statement about leverage, upper tails, or sparse subgroups. The Iris case is equally important because it shows that a visual check can reinforce a finding without being redundant: the plot reveals the geometry behind the ranking of variables.

This role is related to, but weaker than, formal visual inference. A lineup calibrates an observed plot against a null distribution and can support an inferential decision \citep{buja2009statistical,li2024plot}. EDAgent instead preserves a traceable comparison within an exploratory workflow. Its output is a better-scoped interpretation, not a visual test or an inferential guarantee.

\subsection{What the Evaluation Establishes}

The EDA-Bench score distribution indicates that EDAgent can make the relationship among selected statistical findings, charts, and narrative conclusions explicit. The lower score distribution for Critical Data Caveats identifies a different limitation: a report may clearly explain the visual checks it performs while still omitting a relevant issue that was never selected for verification. Improving how the system identifies which findings and potential vulnerabilities require visual follow-up is therefore an important direction for future work.

The adapted VisEval results isolate a narrower component result. VisPlan's observed improvement over Qwen3-8B is concentrated in aesthetics, with smaller changes in alignment, readability, and effectiveness. The qualitative examples show how it operates through more explicit control of palette, spacing, legends, and cross-chart coordination. The results support VisPlan as a compact planning component under this pipeline. They do not show that this architecture is necessary for statistical--visual verification, nor do the tightly clustered model means establish equivalence or superiority among planners.

\subsection{Practical Implications}

The verification workflow is model-agnostic and can be implemented as a lightweight audit layer in other data-agent systems. A practical report should preserve, for every material conclusion, the executed quantity, the reason a visual check was requested, the chart identifier, the observed feature, and the resulting verification outcome. This record makes it easier for an analyst to distinguish what was computed from what was inferred visually and to select follow-up confirmatory analyses. In consequential settings, the appropriate use is decision support: the system can surface leverage, heterogeneity, and distributional caveats, but a human analyst remains responsible for model choice and inference.

\section{Limitations}

The evaluation is not independent. EDA-Bench has no directly evaluated end-to-end comparator, each output is judged once, and GPT-5.1 participates both in the pipeline and in evaluation. Shared model preferences or correlated errors may therefore inflate apparent consistency. The judge receives sampled computational metadata rather than a complete machine-checked ledger of every value, and the ordinal rubric has not been calibrated against expert ratings. The reported proportions and means should be read as descriptive measurements under one fixed protocol.

The evaluation tasks also have limited external validity. EDA-Bench is assembled from a popularity-ranked Kaggle snapshot, while VisPlan is trained from highly voted notebooks. Popularity is not a guarantee of analytical or graphical quality, and both sources may overrepresent tutorial-style, English-language, or presentation-oriented analyses. The open-ended VisEval queries are generated by an LLM, so they may reflect that model's preferences and do not preserve the construct of the original benchmark.

Execution grounding is necessary but insufficient for correctness. Generated code can select an inappropriate variable, compute an unsuitable summary, or run successfully despite a semantic error. The generated chart can encode the wrong subset or scale, and the multimodal model can misread or overstate the rendered pattern. In addition, the current implementation executes code in a local Python subprocess rather than a hardened container. Any deployment on untrusted inputs requires operating-system, file-system, and network isolation beyond the present implementation.

Finally, the framework is limited to static visualizations of tabular data and uses a one-way sequence from statistics to charts to report synthesis. A visual discovery cannot currently trigger a new computation or a redesigned chart. Human evaluation, independent end-to-end comparisons, repeated judging, and an iterative evidence loop are therefore priorities for establishing how well the approach generalizes beyond the current descriptive study.

\section{Conclusion}

EDAgent turns a familiar EDA habit---checking a statistical result against a plot---into an explicit and traceable verification workflow. Executed results establish what was computed, visual verification requests identify distributional features that could affect the interpretation, and rendered charts determine whether the final wording should be retained, qualified, or reconsidered. The contribution is therefore not simply another chart-generation stage; it is a structured account of what a chart examines and how its evidence changes an automated report.

Under the adopted single-judge protocol, EDA-Bench reports most consistently satisfied narrative and evidence-linkage criteria, while VisPlan's observed gain over its base model was concentrated in visual presentation. These findings are promising but deliberately bounded: they do not replace independent baselines, human assessment, or formal statistical validation. EDAgent is intended to help analysts identify summary--structure mismatches and formulate better follow-up questions before confirmatory modeling. The method should ultimately be judged by whether its saved verification records allow an independent analyst to reconstruct why each material interpretation was retained, qualified, or reconsidered.

\section*{Data and Code Availability}

Materials needed to document and reproduce the reported workflow will be supplied to editors and reviewers through an anonymity-preserving channel if requested during peer review. Upon acceptance, the authors will release a public repository containing the implementation code, training-data construction scripts, prompts and evaluation rubrics, the EDA-Bench manifest and screening record, benchmark query text, adapted VisEval tasks, and evaluation scripts. VisPlan adapter weights will also be released subject to upstream model-license terms. Because EDA-Bench uses third-party Kaggle datasets, the repository will provide source identifiers, license metadata, checksums, and reconstruction instructions; original data files will be redistributed only when their licenses permit.

% Add Funding, Acknowledgments, Author Contributions, and Disclosure statements
% only to the non-anonymized manuscript when author details are finalized.

\bibliography{EDAgent_JCGS_v5}

\end{document}